\documentstyle[12pt]{article}
\newcommand{\be}{\begin{eqnarray}}
\newcommand{\ee}{\end{eqnarray}}

\def\lsim{\mathrel{\rlap{\lower4pt\hbox{\hskip1pt$\sim$}}
    \raise1pt\hbox{$<$}}}               
\def\gsim{\mathrel{\rlap{\lower4pt\hbox{\hskip1pt$\sim$}}
    \raise1pt\hbox{$>$}}}               

\begin{document}

\rightline{{\Large Preprint RM3-TH/02-9}}

\vspace{1cm}

\begin{center}

\LARGE{Leading and higher twists in the proton polarized structure function $g_1^p$ at large Bjorken-$x$ \footnote{Proceedings of the IX International Conference on the {\em The Structure of Baryons}, Jefferson Lab (Newport New, USA), March 3-8, 2002, World Scientific Publishing (Singapore), in press.}}

\vspace{1.5cm}

\large{Silvano Simula$^{(*)}$, M. Osipenko$^{(**)}$, G. Ricco$^{(***)}$ and M. Taiuti$^{(***)}$}\\

\vspace{0.5cm}

\normalsize{$^{(*)}$ Istituto Nazionale di Fisica Nucleare, Sezione Roma III\\ Via della Vasca Navale 84, I-00146 Roma, Italy\\ $^{(**)}$Physics Department, Moscow State University, 119899 Moscow, Russia\\ $^{(***)}$ University of Genova and Istituto Nazionale di Fisica Nucleare - Sezione di Genova\\ Via Dodecanneso 33, I-16146, Genova, Italy}

\end{center}

\vspace{1cm}

\begin{abstract}

\noindent Power corrections to the $Q^2$ behavior of the Nachtmann moments of the proton polarized structure function $g_1^p$ are investigated at large Bjorken $x$ by developing a phenomenological fit of both the resonance (including the photon point) and deep inelastic data up to $Q^2 \sim 50 ~ (GeV/c)^2$. The leading twist is treated at $NLO$ in the strong coupling constant and the effects of higher orders of the perturbative series are estimated using soft-gluon resummation techniques. In case of the first moment higher-twist effects are found to be quite small for $Q^2 \gsim 1 ~ (GeV/c)^2$, and the singlet axial charge is determined to be $a_0[10 ~ (GeV/c)^2] = 0.16 \pm 0.09$. In case of higher order moments, which are sensitive to the large-$x$ region, higher-twist effects are significantly reduced by the introduction of soft gluon contributions, but they are still relevant at $Q^2 \sim$ few $(GeV/c)^2$ at variance with the case of the unpolarized transverse structure function of the proton. This finding suggests that spin-dependent correlations among partons may have more impact than spin-independent ones. It is also shown that the parton-hadron local duality is violated in the region of polarized electroproduction of the $\Delta(1232)$ resonance.

\end{abstract}

\newpage

\pagestyle{plain}

\section{Introduction}

\indent The experimental investigation of lepton deep-inelastic scattering ($DIS$) off proton and deuteron targets has provided a wealth of information on parton distributions in the nucleon. In the past few years some selected issues in the kinematical regions corresponding to large values of the Bjorken variable $x$ have attracted a lot of theoretical and phenomenological interest; among them one should mention the occurrence of power corrections associated to {\em dynamical} higher-twist operators measuring the correlations among partons. The extraction of the latter is of particular relevance since the comparison with theoretical predictions either based on lattice $QCD$ simulations or obtained from models of the nucleon structure represents an important test of $QCD$ in its non-perturbative regime.

\indent In Refs. \cite{Ricco} and \cite{SIM00} the world data on the unpolarized nucleon structure functions $F_2^N$ and $F_L^N$ have been used to carry out power correction analyses. In this contribution we summarize the main results \cite{SIM02} of the extension of such a twist analysis to the case of the polarized proton structure function $g_1^p$, performed in terms of Nachtmann moments. The latter however require the knowledge of the polarized structure function $g_1^p$ in the whole $x$-range for fixed values of $Q^2$, and therefore a new parameterization of $g_1^p$, which describes the $DIS$ proton data up to $Q^2 \sim 50 ~ (GeV/c)^2$ and includes a phenomenological Breit-Wigner ansatz able to reproduce the existing electroproduction data in the proton-resonance regions, has been developed \cite{SIM02}. The interpolation formula for $g_1^p$ has been successfully extended down to the photon point, showing that it nicely reproduces the very recent data \cite{Mainz} on the energy dependence of the asymmetry of the transverse photoproduction cross section as well as the experimental value of the proton Drell-Hearn-Gerasimov ($DHG$) sum rule. According to our parameterization of $g_1^p$ the generalized $DHG$ sum rule is predicted to have a zero-crossing point at $Q^2 = 0.16 \pm 0.04 ~ (GeV/c)^2$. Finally, the $Q^2$ behavior of low-order polarized Nachtmann moments has been obtained in the $Q^2$-range between $0.5$ and $50 ~ (GeV/c)^2$. 

\section{Parton-hadron local duality in $g_1^p$}

\indent An important feature of the results obtained in Ref. \cite{SIM02} is that the resonant contribution to the polarized Nachtmann moments is negative for $Q^2 \sim$ few $(GeV/c)^2$. This is mainly due to the well established fact \cite{Burkert} that the proton transverse asymmetry $A_1^p$ in the $\Delta(1232)$-resonance regions is negative up to $Q^2 \sim 3 \div 4 ~ (GeV/c)^2$. Thus, for $Q^2 \sim$ few $(GeV/c)^2$ the resonant contribution to the polarized proton structure function is opposite in sign with respect to the unpolarized case (see Ref. \cite{Ricco}).

\indent It is therefore legitimate to ask ourselves whether the parton-hadron local duality, observed empirically \cite{BG} in the unpolarized transverse structure function of the proton, holds as well in the polarized case. To this end we have generated pseudo-data in the resonance regions and in the $DIS$ regime via our interpolation formula for $g_1^p$. Our results clearly shows that: ~ i) at values of $Q^2$ as low as $\sim 0.5 ~ (GeV/c)^2$ there is no evidence at all of an occurrence of the local duality, as in the case of the unpolarized transverse structure function of the proton \cite{duality}, and ~ ii)  in the kinematical regions where the $\Delta(1232)$ resonance is prominently produced, the local duality breaks down at least for $Q^2$ up to few $(GeV/c)^2$, while in the higher resonance regions for $Q^2 \gsim 1 ~ (GeV/c)^2$ it is not excluded by our parameterization. Note that in the unpolarized case the onset of the local duality occurs \cite{BG,duality} at $Q^2 \simeq 1 \div 2 ~ (GeV/c)^2$, including also the $\Delta(1232)$ resonance regions. It should be mentioned that the usefulness of the concept of parton-hadron local duality relies mainly on the possibility to address the $DIS$ curve at large $x$ through measurements at low $Q^2$ in the resonance regions. It is therefore clear that the breakdown of the local duality in the region of the $\Delta(1232)$ resonance forbid us to get information from duality on the behavior of the scaling curve at the highest values of $x$.

\section{Twist analysis of the polarized Nachtmann moments}

\indent In this Section we present the power correction analysis of the polarized Nachtmann moments, $M_n^{(1)}(Q^2)$, obtained in Ref. \cite{SIM02}. The leading twist, $\mu_n^{(1)}(Q^2)$, is treated both at next-to-leading ($NLO$) order and beyond any fixed order by adopting available soft gluon resummation ($SGR$) techniques. As for the power corrections, a phenomenological ansatz is considered, viz.
 \be
       M_n^{(1)}(Q^2) =\mu_n^{(1)}(Q^2) + a_n^{(4)} {\mu^2 \over Q^2} 
       \left[ {\alpha_s(Q^2) \over \alpha_s(\mu^2)} \right]^{\gamma_n^{(4)}} +  
       a_n^{(6)}  \left( {\mu^2 \over Q^2} \right)^2 \left[ {\alpha_s(Q^2) \over 
       \alpha_s(\mu^2)} \right]^{\gamma_n^{(6)}}
       \label{eq:M1n}
 \ee
where the logarithmic $pQCD$ evolution of the twist-4 (twist-6) contribution is accounted for by an effective anomalous dimension $\gamma_n^{(4)}$ ($\gamma_n^{(6)}$) and the parameter $a_n^{(4)}$ ($a_n^{(6)}$) represents the overall strength of the twist-4 (twist-6) term at the renormalization scale $\mu^2$, chosen to be equal to $\mu^2 = 1 ~ (GeV/c)^2$. In order to fix the running of the coupling constant $\alpha_s(Q^2)$, the updated $PDG$ value $\alpha_s(M_Z^2) = 0.118$ is adopted.

\indent In case of the first Nachtmann moment ($n = 1$) the corresponding leading twist term, $\delta\mu_1^{(1)}(Q^2)$, does not receive any correction from $SGR$ and therefore at $NLO$ it reads as 
 \be
       \mu_1^{(1)}(Q^2) = {<e^2> \over 2} \left[ \Delta q^{NS} + a_0(Q^2) 
       \right] \left[ 1 - {\alpha_s(Q^2) \over \pi} \right] 
       \label{eq:mu1_first}
 \ee
The non-singlet moment $\Delta q^{NS}$ is taken fixed at the value $\Delta q^{NS} = 1.095$, deduced from the experimental values of the triplet and octet axial coupling constants, with the latter obtained under the assumption of $SU(3)$-flavor symmetry. The values of the singlet axial charge $a_0(\mu^2)$ and of the four higher-twist quantities $a_1^{(4)}$, $\gamma_1^{(4)}$, $a_1^{(6)}$ and $\gamma_1^{(6)}$ are determined by fitting our pseudo-data, adopting the least-$\chi^2$ procedure in the $Q^2$-range between $0.5$ and $50 ~ (GeV/c)^2$. It turns out that the total contribution of the higher twists is tiny for $Q^2 \gsim 1 ~ (GeV/c)^2$, but it is comparable with the leading twist already at $Q^2 \simeq 0.5 ~ (GeV/c)^2$. Since the first moment basically corresponds to the  area under the structure function $g_1^p$ (as it is the case of the second moment of the unpolarized structure function $F_2^p$), the dominance of the leading twist in $M_1^{(1)}(Q^2)$, occurring for $Q^2 \gsim 1 ~ (GeV/c)^2$, reflects only the concept of global duality and {\em not} that of local duality (cf. Ref. \cite{duality}). In our analysis, where the leading and the higher twists are simultaneously extracted, the singlet axial charge (in the $AB$ scheme) is determined to be $a_0(10 ~ GeV^2) = 0.16 \pm 0.09$, which nicely agrees with many recent estimates appeared in the literature. Our value of $a_0$ is therefore significantly below the naive quark-model expectation (i.e. compatible with the well known "proton spin crisis"), but it does not exclude completely a singlet axial charge as large as $\simeq 0.25$.

\indent In case of higher-order moments ($n \geq 3$) both the $NLO$ approximation and the $SGR$ approach have been considered for the leading twist. The comparison of the  corresponding twist analyses shows \cite{SIM02} that, except for the third moment, the contribution of the twist-2 is enhanced by soft gluon effects, while the total higher-twist term decreases significantly after the resummation of soft gluons. Thus, as already observed \cite{SIM00} in the unpolarized case, also in the polarized one it is mandatory to go beyond the $NLO$ approximation and to include soft gluon effects in order to achieve a safer extraction of higher twists at large $x$, particularly for $Q^2 \sim$ few $(GeV/c)^2$.

\indent Finally, the twist decomposition of the polarized Nachtmann moments has been compared with the corresponding one of the unpolarized (transverse) Nachtmann moments obtained in Ref. \cite{SIM00} adopting the same $SGR$ technique. It turns out  \cite{SIM02} that the extracted higher-twist contribution appears to be a larger fraction of the leading twist in case of the polarized moments. This findings suggests that spin-dependent multiparton correlations may have more impact than spin-independent ones.

\end{document}